\documentclass[journal, 10pt]{IEEEtran}
\usepackage{graphicx}
\usepackage{amsmath}
\usepackage{amssymb}
\usepackage{subfigure}
\usepackage[noadjust]{cite}
\usepackage[warn]{textcomp}
\usepackage{multirow}
\usepackage{booktabs}
\usepackage{tabularx}
\usepackage{color,soul}
\usepackage[detect-all=true]{siunitx}
\usepackage{verbatim}
\usepackage{floatrow}
\usepackage[utf8]{inputenc}
\usepackage{array}
\usepackage{makecell}

\usepackage{caption}
\usepackage{float}
\usepackage{siunitx}
\restylefloat{table}
\usepackage{amssymb}
\usepackage[tableposition=top]{caption}
\usepackage{dblfloatfix}
\usepackage{float}
\floatstyle{plaintop}
\restylefloat{table}
\usepackage{authblk}
\usepackage{booktabs}
\usepackage{multirow} 
\usepackage{multicol}
\usepackage{rotating} 
\usepackage{amsmath}
\usepackage[font=small,labelfont=bf]{caption}
\usepackage{lettrine} 
\usepackage{lipsum}
\usepackage{setspace}
\newcommand{\ie}{\textit{i.e.}}

\newcommand{\eg}{\textit{e.g.}}

\begin{document}
\IEEEoverridecommandlockouts

{\setstretch{1}
\title{\huge{Exploiting Dual-Gate Ambipolar CNFETs for Scalable Machine Learning Classification}
   \vspace{0.15in}
}
\author[1]{\hspace{-5pt}Farid Kenarangi}
\author[2]{Xuan Hu}
\author[3]{Yihan Liu}
\author[3]{Jean Anne C. Incorvia}
\author[2]{Joseph S. Friedman}
\author[1]{Inna Partin-Vaisband}
\affil[1]{University of Illinois at Chicago}
\affil[2]{University of Texas at Dallas}
\affil[3]{University of Texas at Austin}

\maketitle

\begin{abstract} 
Ambipolar carbon nanotube based field-effect transistors (AP-CNFETs) exhibit unique electrical characteristics, such as tri-state operation and bi-directionality, enabling systems with complex and reconfigurable computing. In this paper, AP-CNFETs are used to design a mixed-signal machine learning (ML) classifier. The classifier is designed in SPICE with feature size of \SI{15}{\nano\meter} and operates at \SI{250}{\mega\hertz}. The system is demonstrated based on MNIST digit dataset, yielding 90\% accuracy and no accuracy degradation as compared with the classification of this dataset in Python. The system also exhibits lower power consumption and smaller physical size as compared with the state-of-the-art CMOS and memristor based mixed-signal classifiers.

\end{abstract}
\section{INTRODUCTION}
\lettrine{P}{ower} consumption and physical size of integrated circuits (ICs) is an increasing concern in many emerging ML applications, such as, autonomous vehicles, security systems, and Internet of Things (IoT). 
Existing state-of-the-art architectures for digital classification, are highly accurate and can provide high throughput \cite{lee2018unpu}. These classifiers, however, exhibit high power consumption and occupy a relatively large area to accommodate complex ML models. Alternatively, mixed-signal classifiers have been demonstrated to exhibit orders of magnitude reduction in power and area as compared to digital classifiers with prediction accuracy approaching the accuracy of digital classifiers \cite{zhang2017memory,wang2017low,bankman2018always,kenarangi2019single,kang2009chip,gonugondla2018variation,kang2018multi,amaravati201855}. 
While significant advances have been made at the ML circuit and architecture levels (\eg, in-SRAM processing \cite{zhang2017memory}, comparator based computing \cite{wang2017low}, and switched-capacitor neurons \cite{bankman2018always}), the lack of robust, ML-specific transistors is a primary concern in ML training and inference with all conventional CMOS technologies. To efficiently increase the density and power efficiency of modern ML ICs while enabling complex computing, emerging technologies should be considered. 

While the non-volatility of memristors has proven quite attractive for storing the weights required for feature-weight multiplication \cite{hu2018memristor,yu2015scaling,agarwal2016resistive,krestinskaya2018learning}, field-effect transistors provide several advantages over memristors for ML. First, transistors provide a broader range of linear tuning of resistance, thereby better matching ML models. Second, transistors are not subject to the deleterious aging that deteriorates memristor behavior over time. Finally, the connectivity between feature-weight multiplication layers requires electrical signal gain, which cannot be provided by memristors; transistors can be used for such interlayer connections, thereby enabling a monolithic integrated circuit that can be fabricated efficiently. Of particular interest for on-chip classification are ambipolar devices. Owing to unique electrical characteristics, as described in Section \ref{sec:background}, ambipolar devices are expected to provide efficient on-chip training and inference solutions and reduce design and routing complexity of ML circuits.
\begin{figure}[t]
	\centering
	\includegraphics[width=0.8\columnwidth]{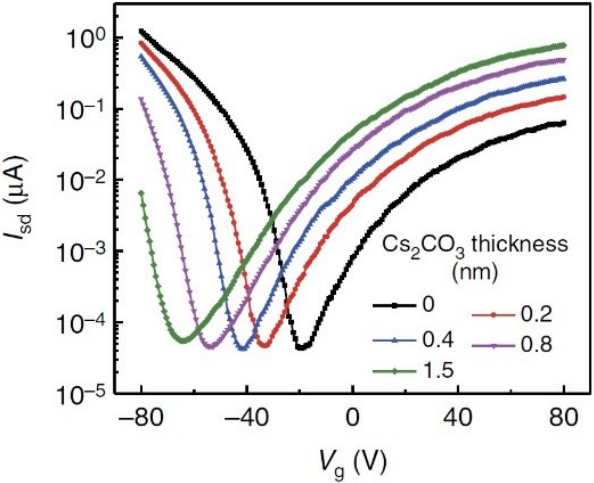} 
	\caption{Forward-transfer characteristics (bottom gate voltage varies from -\SI{80}{\volt} to \SI{80}{\volt}) evolution of a BP FET measured at $V_{sd}=$\SI{100}{\milli\volt} in logarithmic scale with increasing $Cs_2CO_3$ thickness from 0 to \SI{1.5}{\nano\meter} \cite{xiang2015surface}.}
	\label{fig:Fig_1}
\end{figure}

A carbon nanotube ambipolar device has been reported as a potential candidate for controllable ambipolar devices because of its satisfying carrier mobility and its symmetric and good subthreshold ambipolar electrical performance \cite{lin2005high}.  Based on the dual-gate CNT device’s electrical performance, a library of static ambipolar CNT dual-gate devices based on generalized NOR-NAND-AOI-OAI primitives, which efficiently implements XOR-based functions, has been reported, indicating a performance improvement of $\times$7, a 57\% reduction in power consumption, and a $\times$20 improvement in energy-delay product over the CMOS library \cite{ben2011efficient}.

Besides CNTs, some 2D semiconductors such as $MoS_2$, $WS_2$, $WSe_2$ and black phosphorus (BP) are also reported as ambipolar semiconductors at room temperature. Surface transfer doping and using different source and drain contact metal are reported as two effective way to modulate its ambipolar characteristics to move the subthreshold curve’s symmetric point to $V_{bottomgate}=0$ and reduce the difference between the n-branch and p-branch saturation current. Fig. \ref{fig:Fig_1} indicates that undoped multiplayer BP FETs’ saturation current and mobility in the n-branch are much lower than that of the p-branch. However, $Cs_2CO_3$ layers deposited over BP serve as an efficient n-type surface dopant to improve the electron transport in the BP devices, thereby inducing either a more balanced ambipolar or even-transport-dominated FET behavior \cite{xiang2015surface}. Also, by using Ni as the source metal and Pd as the drain contact, the experimental transfer characteristics curve of a multilayer $WSe_2$ FET indicates this configuration allows for ambipolar characteristics with both the electron and hole conduction current levels being similar \cite{das2013wse2}.

Existing results exploit the switching characteristics of the AP-CNFETs for enhancing digital circuits \cite{ben2011efficient,o2007cntfet}. In this work, we  repurpose the AP-CNFET device for neuromorphic computing. Owing to the dual gate structure, AP-CNFET significantly increases the overall density of analog ML ICs, simplifies routing, and reduces power consumption. To the best of the authors knowledge, the AP-CNFET based ML framework is the first to demonstrate a multiplication-accumulation (MAC) operation with single-device-single-wire configuration. 
Note that in CMOS classifiers at least two sensing lines are required to separately accumulate results of multiplication with positive and negative weights. Furthermore, additional circuitry is required to process the signals from the individual sensing lines into a final prediction. Alternatively, in memristor based classifiers, a crossbar architecture is typically used with a single sensing line per class. With this configuration, only positive (or negative) feature weights are, however, utilized. Thus, additional non-linear thresholding circuits are required for extracting the final decision. 

The rest of the paper is organized as follows. The device background and electrical characteristics of AP-CNFET device are presented in Section \ref{sec:background}. The proposed scheme for utilizing AP-CNFETs for on-chip classification is described in Section \ref{sec:Classifier}. The classifier is evaluated based on classification of commonly used Modified National Institute of Standards and Technology (MNIST) dataset, as explained in Section \ref{sec:results_sec}. Finally, the paper is concluded in Section \ref{sec:Conc}. 
\section{BACKGROUND} \label{sec:background}
\subsection{AP-CNFET Device}
Depending on the gate voltage, an ambipolar device allows both electrons and holes to flow from source to drain because of its narrow Schottky barrier width, as small as a couple of nanometers, between metal contacts and the channel \cite{lin2005high}. Fig. \ref{fig:Fig_2} shows the schematic of a top-bottom dual-gate ambipolar device. In contrast to a normal single-gate-control transistor, the bottom gate plays an important role in determining the type of majority carrier and on current. This dual-gate device’s electrical characteristic can be understood by the schematic band diagrams shown in Fig. \ref{fig:Fig_3}. For a sufficiently negative (positive) bottom gate voltage, the Schottky barrier is thinned enough to allow for holes (electrons) tunneling from the source contact into the channel to the drain. By tuning the top gate voltage more positive (negative) to alter the barrier height for carrier transport across the channel, the top gate can switch between the ON and OFF operating states.
\begin{figure}[t]
	\centering
	\includegraphics[width=0.8\columnwidth]{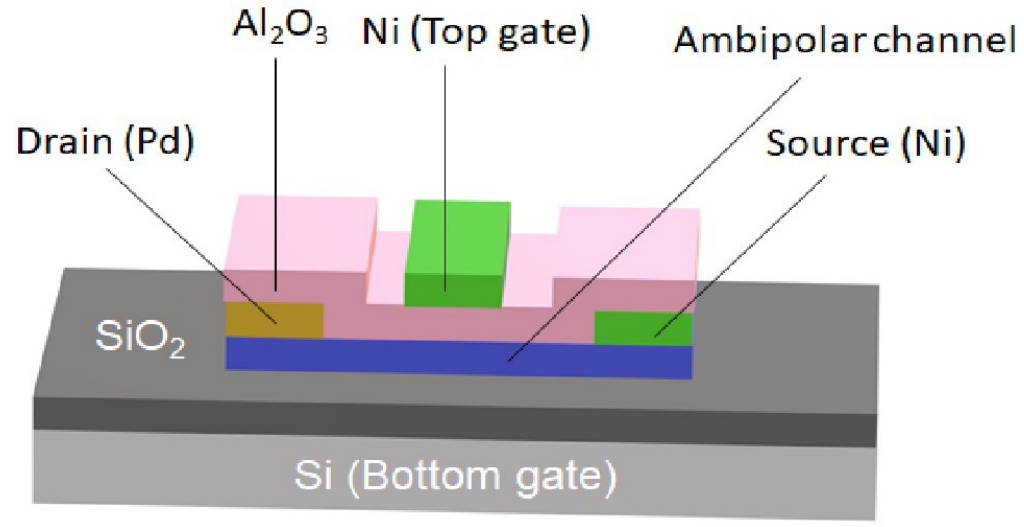} 
	\caption{Schematic of a top-bottom dual-gate ambipolar device.}
	\label{fig:Fig_2}
\end{figure}
\begin{figure}[t]
	\centering
	\includegraphics[width=0.8\columnwidth]{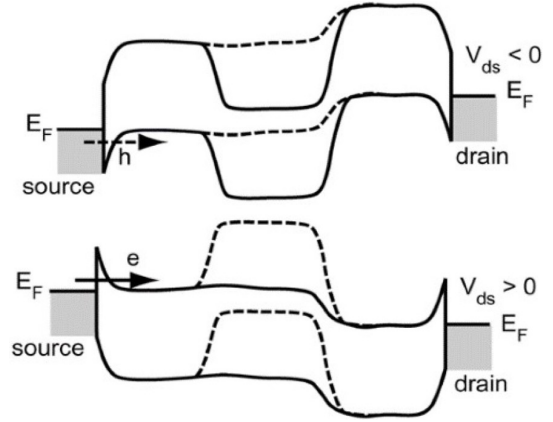} 
	\caption{Schematic band diagrams of a dual-gate for bottom gate voltage $< 0$ and $> 0$. The solid and dashed lines at the middle region show how the top gate voltage switches the operating state by allowing or stopping the current flow \cite{lin2005high}.}
	\label{fig:Fig_3}
\end{figure}
\subsection{Electrical Characteristics of the AP-CNFET}
\begin{figure*}[t]
	\centering
	\includegraphics[width=0.9\columnwidth]{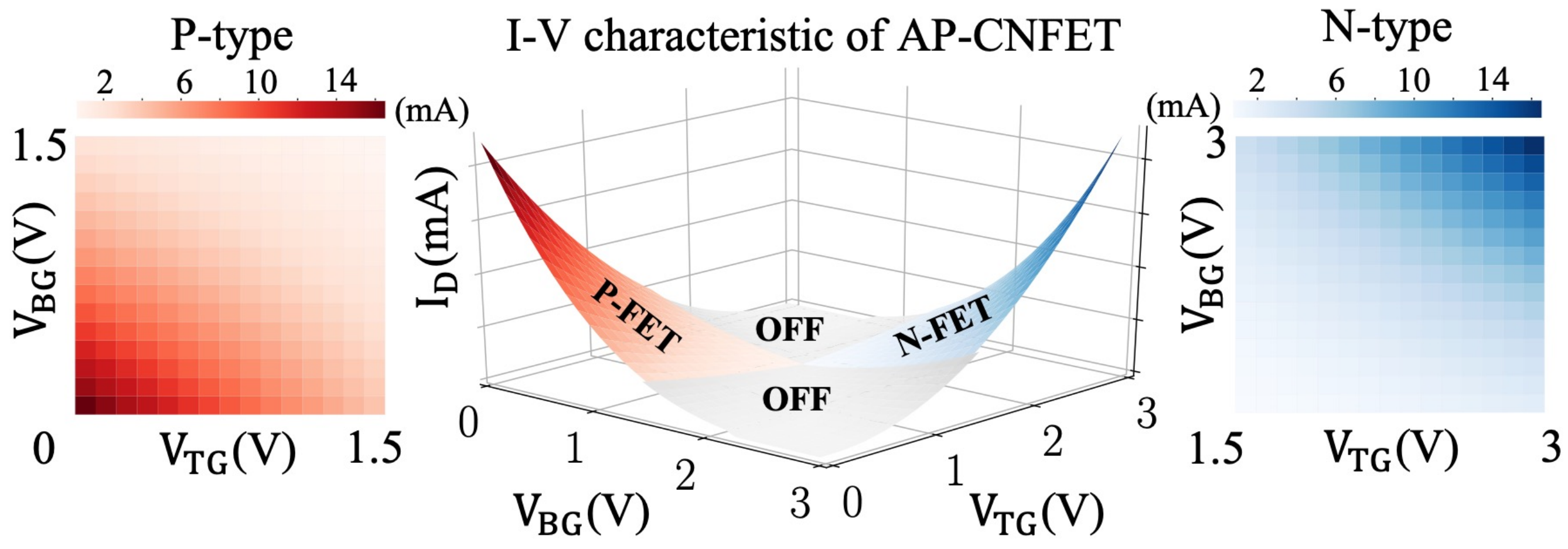} 
	\caption{IV-characteristics of the AP-CNFETs.}
	\label{fig:IV}
\end{figure*}
As compared with a conventional MOSFET, an AP-CNFETs exhibits two unique characteristics as explained below.  

\textbf{Tri-State Operation.}  
Due to ambipolarity of an AP-CNFET, the majority carriers in the device can be either electrons or holes, depending upon the gate biases. The same device can, therefore, operate as n-type (majority carriers are electrons) or p-type (majority carriers are holes) FET. The I-V characteristics (\ie, drain current $I_D$ versus top gate voltage $V_{TG}$ and bottom gate voltage $V_{BG}$) of a single AP-CNFET are shown in Fig. \ref{fig:IV}, exhibiting p-type (red), n-type (blue), and OFF (gray) operational regions. 
In typical mixed-signal ML classifiers, positive and negative classification decisions are separately accumulated on the individual sensing lines \cite{zhang2017memory,kang2009chip,kenarangi2019leveraging,kenarangi2019single}. Thus, at least two wires are required for a single MAC operation. Alternatively, with the proposed structure, the unique tri-state operation of AP-CNFETs is leveraged for merging the sensing lines, significantly reducing the routing complexity and area overhead, as described in Section \ref{sec:Classifier}.

\textbf{Bi-directionality.} Another unique characteristic, is the bi-directionality of the device \cite{hu2017transient,hu2017closed}. The drain and source terminals are determined based on the potential difference across the device. The terminals with higher and lower potentials act as, respectively, the drain and source nodes. Exploiting the bi-directionality of the AP-CNFETs, highly reconfigurable systems can be targeted. One approach is using bi-directionality for on-chip training, where the current direction within the individual AP-CNFETs is adjusted during each training iteration. By interchanging the drain and source terminals, similar current values yet in opposite directions can be generated, giving rise to fundamentally new, efficient, and reconfigurable implementation of ML algorithms. 

\section{LEVERAGING AP-CNFET FOR MACHINE LEARNING CLASSIFICATION} \label{sec:Classifier}
In this section, AP-CNFET based ML classification is demonstrated. The ML background is provided in Section \ref{sec:ML_background}. The circuit level operation principles are explained in Section \ref{sec:Circuit}. 
\subsection{ML Background} \label{sec:ML_background}
Owing to its dual gate structure and electrical properties, AP-CNFET is a natural choice for embedded AI. The superiority of the AP-CNFET based learning over the traditional approaches is demonstrated in this paper based on the linear ML classification problem. Linear predictors are commonly preferred for on-chip classification due to their simplicity, satisfactory performance in categorizing linearly separable data, and low design complexity and hardware costs. Note that the proposed scheme is robust and can be utilized with more complex systems, such as non-linear ML classifiers and deep neural networks. With a multivariate linear classifier, the system response $Z$ is an accumulated dot product of $N$ input features $x=({x_1}, {x_2}, ...,{x_N})$ and model weights $w=({w_1},
{w_2}, ...,{w_N})$,
\begin{equation} \label{eq:Z}
Z=\sum_{i=1}^{N}w_{i}\cdot x_{i}, \quad Z\in \mathbb R.
\end{equation}
Logistic regression (LR) algorithm is utilized in this paper to train the classifier based on gradient descent algorithm \cite{nelder2004generalized}. The model weights, $w$, are determined during supervised training by minimizing the prediction error between the system response, $Z$, and labeled training dataset. In inference, the probability threshold is used for predicting system response to unseen input data, exhibiting a simple on-chip implementation, 

\begin{equation} \label{eq:decision}
\resizebox{.91\hsize}{!}{$\hat{y}=\text{sign}(Z)=\text{sign}(\sum_{i=1}^{N}w_{i} \cdot x_{i})
	=\left\{
	\begin{array}{ll}
	1,  \quad  &\:Z\ge Z_{th}\\
	-1, \quad  &\:Z <Z_{th}.\\
	\end{array}
	\right.$}
\end{equation}

The described logistic regressor with the probability threshold of $Z_{th}=0$ is referred to as logistic classifier. 
The accuracy of the proposed logistic classifier is evaluated as the percentage of all the correct predictions out of the total number of the unseen test data points. 

\subsection{Design of AP-CNFET Classifier} \label{sec:Circuit}
\begin{figure}[t]
	\centering
	\includegraphics[width=1\columnwidth]{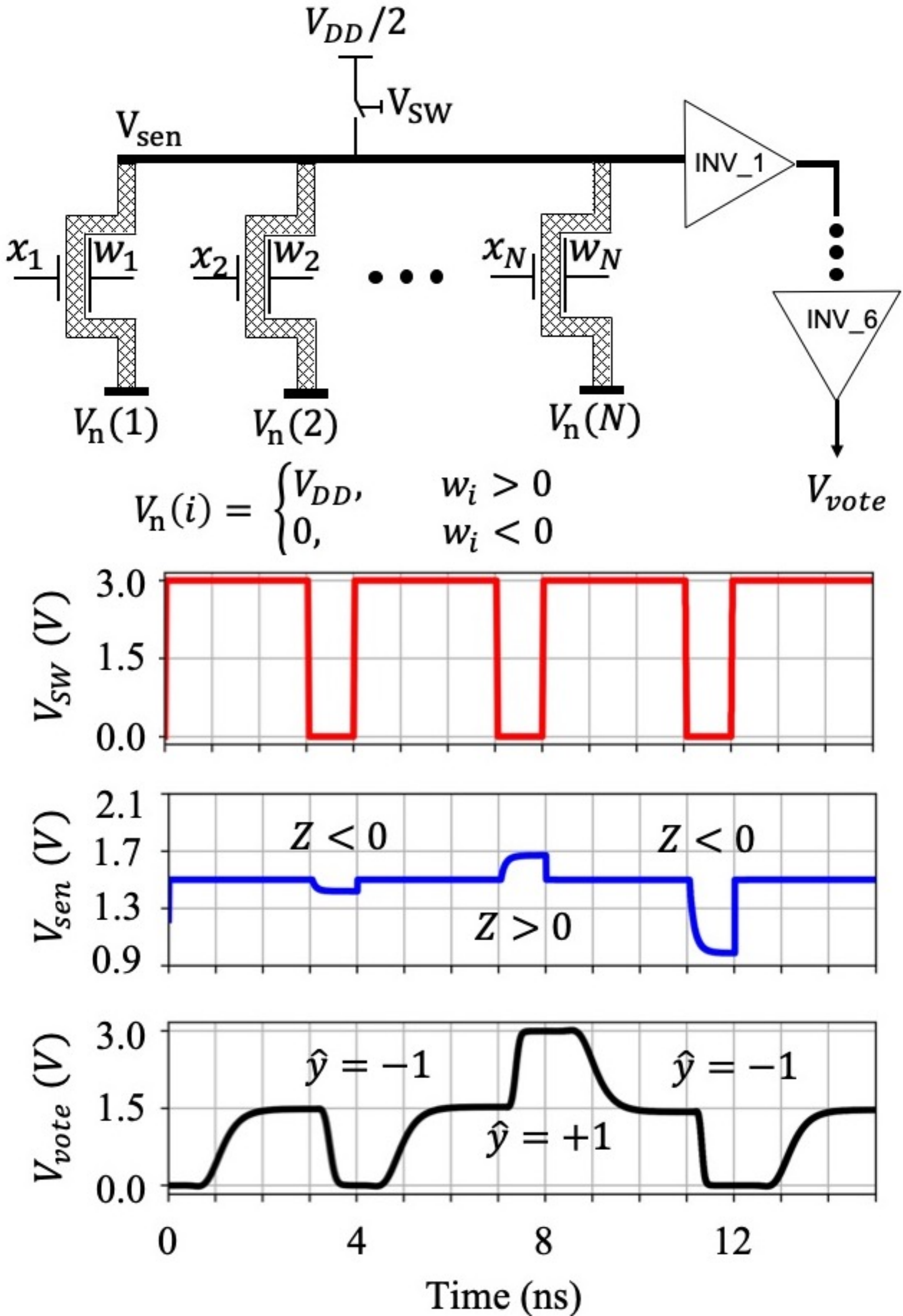} 
	\caption{The overall schematics of the proposed $N$-dimensional, binary classifier. The top and bottom gates of the AP-CNFETs are connected to, respectively, the corresponding features, $x_i$, and feature weights, $w_i$. To encode the weight sign, the individual AP-CNFETs are biased as either p-type ($w_i > 0$) or n-type ($w_i < 0$) states. The circuit operates in two stages: precharge and classification (see the red $V_{SW}$ waveform). During the precharge state, the sensing line is charged to $V_{DD}/2$. The classification decision is made during the classification phase based on the line voltage level as compared to $V_{DD}/2$ (see the blue $V_{sen}$ waveform). The final decision between two classes is made based on the output of the buffers chain (see the black $V_{vote}$ waveform).}
	\label{fig:circuit}
\end{figure}
The overall schematics of the classifier is shown in Fig. \ref{fig:circuit}. The circuit is designed to perform a $N$-feature binary classification. A single sensing line is used for storing the classification results produced by $N$ AP-CNFETs attached to the line. The top and bottom gates of the transistors are connected to, respectively, the corresponding features and feature weights. Consequently the current through each transistor is proportional to the feature-weight product. The tri-state attribute of the AP-CNFETs is exploited to encode the sign of the individual products, facilitating the accumulation of the multiplication results on a single sensing line. To encode the sign, the individual AP-CNFETs are biased as either p-type (for $w_i>0$) or n-type (for $w_i<0$) states. As a result, a certain amount of charge, as determined by the feature-weight product, is injected into (by p-type) or removed from (by n-type) the sensing line. To enable the tri-state operation, each device is connected between the common sensing line and either the $V_{DD}$ (p-type) or ground (n-type) supply.

The circuit operates in two stages: precharge and classification. During the precharge state, the sensing line is charged to $V_{DD}/2$ and transistors are gated with appropriate top gate biases (\ie, $V_{DD}$ for p-type and ground for n-type). During classification, the transistors are biased with their corresponding feature and feature weight values.
As a result, the line is further charged (with p-type FETs) or discharged (with n-type FETs). Depending on the signs of the individual feature-weight products, the sensing line accumulates or dissipates a certain amount of charge, saturating at, respectively, higher ($>V_{DD}/2$) or lower ($<V_{DD}/2$) voltage $V_{sen}$. Consequently, the classification decision is determined as,
\begin{equation} 
\resizebox{.55\hsize}{!}{$\hat{y}
	=\left\{
	\begin{array}{ll}
	+1,  \quad  &\:V_{sen} > V_{DD}/2\\
	-1, \quad  &\:V_{sen} < V_{DD}/2.\\
	\end{array}
	\right.$}
\end{equation}

To extract the final decision, a non-inverting buffer is connected to the sensing line, as shown in Fig. \ref{fig:circuit}. For $V_{sen} > V_{DD}/2$, the output of the sensing line is forced to $V_{vote} = V_{DD}$. The line is forced to $V_{vote} = 0$ for $V_{sen} < V_{DD}/2$. Note that using a single sensing line eliminates a conventional classification stage where the voltage on the positive and negative sensing lines is compared to determine the final binary classification result. The proposed circuit is also more resilient to process, voltage, and temperature (PVT) variations as compared with state-of-the-art mixed-signal classifiers, since it only uses inverters to extract the final decision. Alternatively, in state-of-the-art classifiers (\eg, \cite{zhang2017memory,wang2017low}), comparators, are utilized to extract the final decision. The correct operation of such comparators highly relies on symmetry of the circuit, exhibiting significant offset under PVT variations. To mitigate the sensitivity to PVT variations, additional compensating rows are utilized in the existing ML classifiers. \cite{zhang2017memory}. The small power consumption and area of these ML classifiers is, therefore, traded off for higher classification accuracy.
	
Linearity of the individual AP-CNFETS is critical for correct classification of the input features. Albeit the semi-linear dependence of output current on the gate biases across the full voltage range of operation (see Fig. \ref{fig:IV}), AP-CNFET exhibits no degradation in classification accuracy as compared to classification accuracy in Python, as presented in Section \ref{sec:results_sec}. Alternatively, utilization of the wide bias region allows for quantization of features and feature weights with larger quantization step, increasing the resilience of the circuit to PVT variations. In this paper, five-bit resolution is considered for quantizing features and feature weights with a \SI{40}{\milli\volt} step size. To quantize the features and feature weights, resistive voltage dividers are used. While feature weight connections are set to fixed values, multiplexer (MUX) units are used to update the features within each classification period. Note that to support reconfigurable feature weights, memory units (for storing the weights) along with multiplexer units (for selecting the desired weights) can be utilized. While power overhead is negligible with this approach, the overall area is expected to be increased by a factor of four.

\section{SYSTEM DEMONSTRATION AND RESULTS} \label{sec:results_sec}
The classifier is designed in SPICE and evaluated based on a commonly used MNIST dataset. The dataset and preprocessing steps are described in Section \ref{sec:dataset}. The overall system and SPICE simulation results are presented in Section. \ref{sec:results}.

\subsection{Dataset and Preprocessing Steps} \label{sec:dataset}
MNIST is a large dataset of digit images commonly used for evaluating the effectiveness of ML ICs. MNIST contains images of 70,000 handwritten digits, ranging between
0 and 9. Each digit comprises 784 (28 $\times$ 28) image pixels. The default training and test datasets comprise, respectively, 60,000 and 10,000 digits. Out of the 60,000 training observations, 45,000 and 15,000 are used for, respectively, the training and validation of the proposed system. 

One versus one classification scheme \cite{aly2005survey}, is used to discriminate the 10-class MNIST dataset. A $K$-class, one versus one classifier is designed with $K(K-1)/2$ binary classifiers for pairwise discrimination of the digits. Each binary classifier votes for a single class and the class with highest number of votes is selected as the final classification decision. Utilizing the full set of features (\ie, 784 features), accuracy of 94\% can be achieved on MNIST test set with $10(10-1)/2=45$ binary logistic classifiers. Alternatively, a subset of the 784 features (\ie, 23 features on average) is used in this paper, trading off the performance (less than 4.6\% accuracy degradation) for power and area efficiency ($\times (784/23)=\times 34$ less transistors).  
\begin{figure}[t]
	\centering
	\includegraphics[width=0.7\columnwidth]{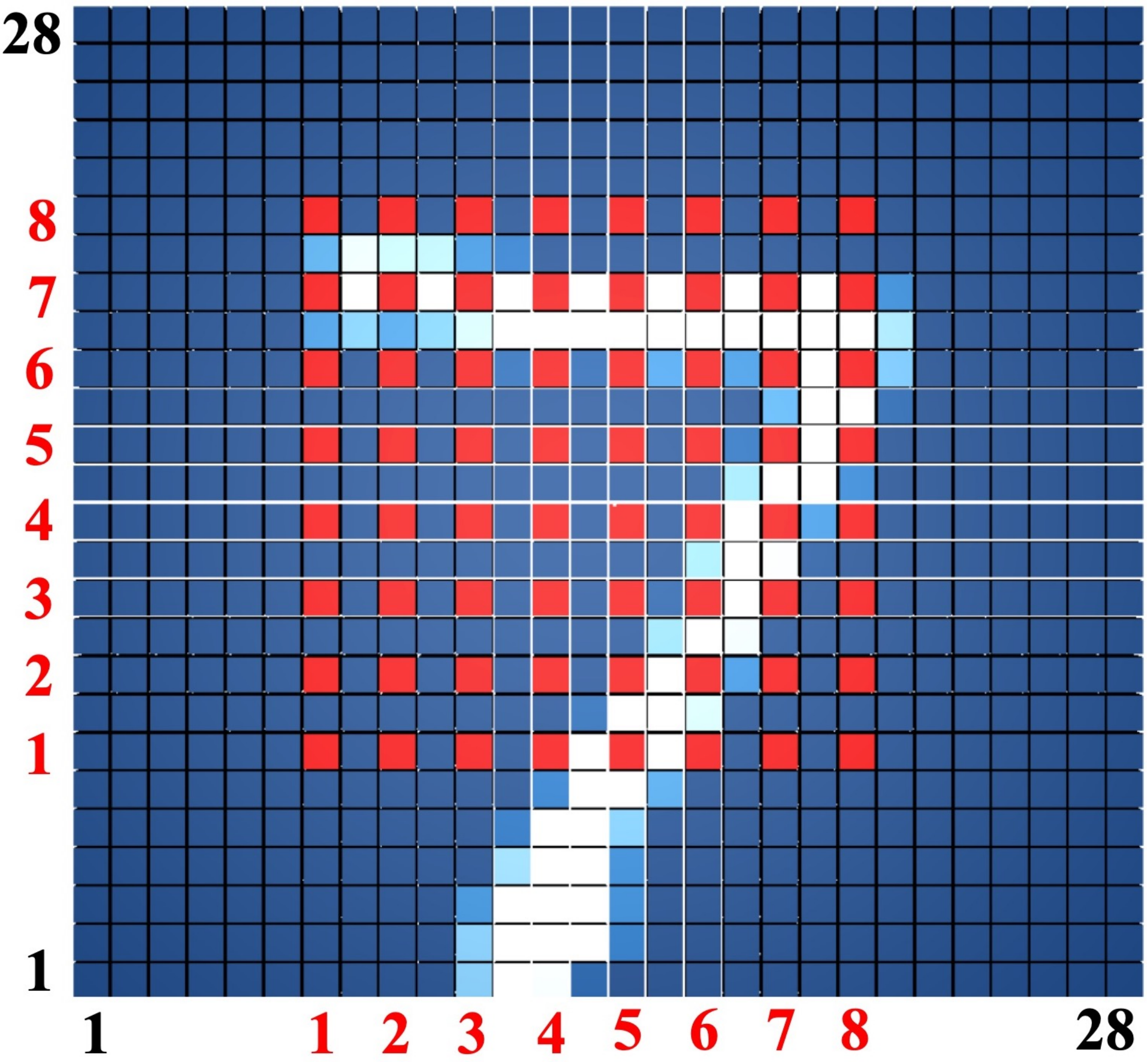} 
	\caption{An example of MNIST digit image. Out of the total 784 features, 720 features are dumped during the downsampling stage. The selected 64 (8 $\times$ 8) features are shown by red squares.}
	\label{fig:selection}
\end{figure}
\begin{figure}[t]
	\centering
	\includegraphics[width=0.8\columnwidth]{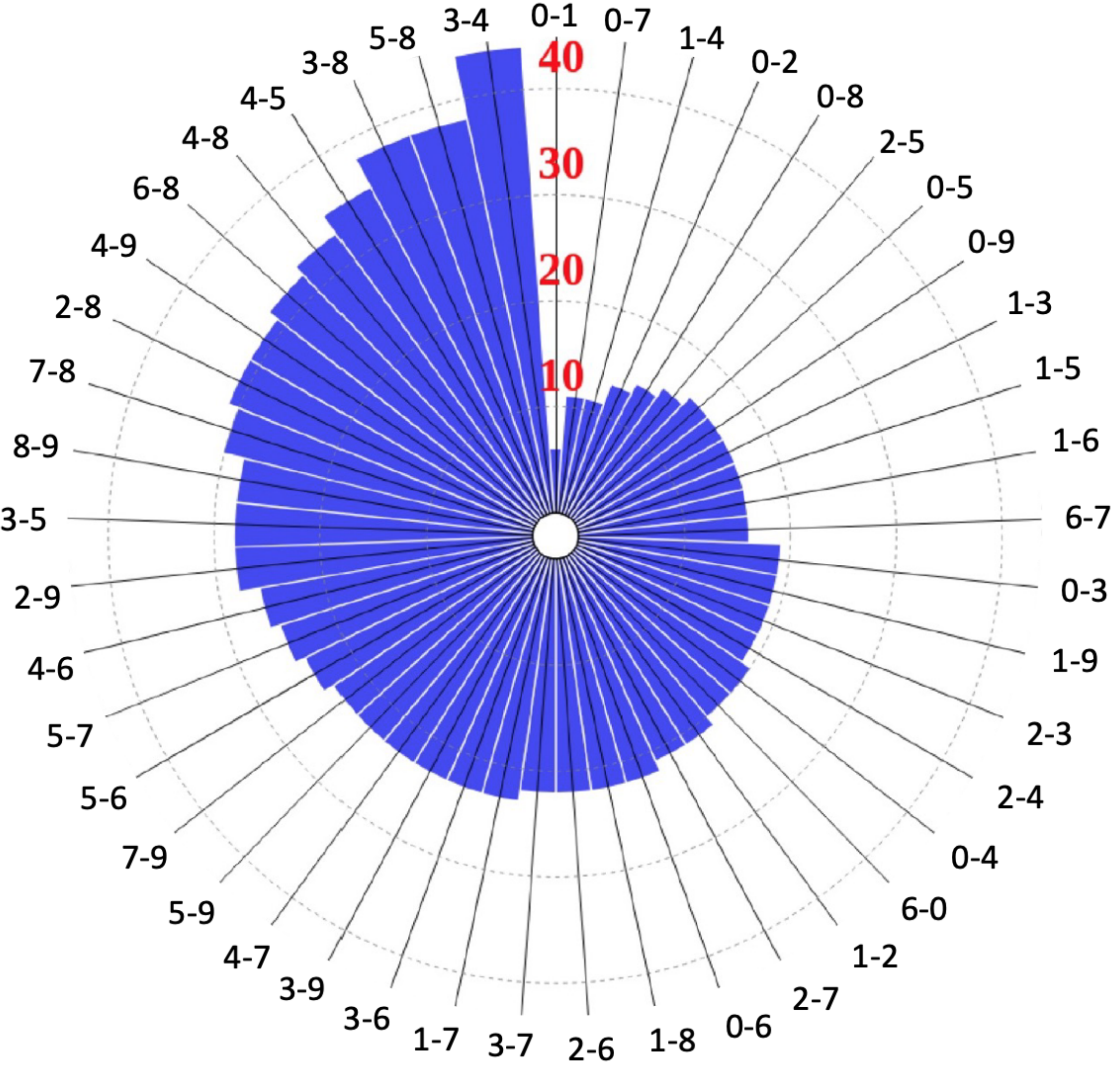} 
	\caption{The selected number of features for each binary classifier (\ie, i-j classifier).}
	\label{fig:sbs}
\end{figure}
\begin{figure}[t]
	\centering
	\includegraphics[width=1\columnwidth]{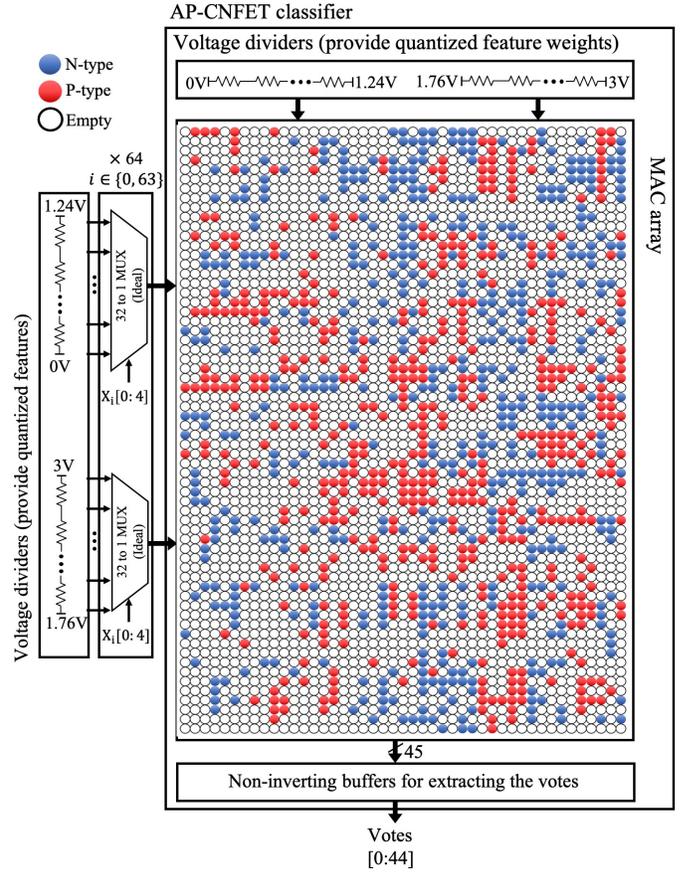} 
	\caption{A schematic diagram of the proposed AP-CNFET based 10-class classifier. An array of $64\times 45$ AP-CNFETs is used to perform the feature-weight products. The transistors biased in n-type and p-type regions are shown by, respectively, the blues and red circles. The transistors that have been removed as a result of SBS feature reduction are shown by white circles.}
	\label{fig:overall_system}
\end{figure}
\begin{figure*}[t]
	\centering
	\includegraphics[width=1\columnwidth]{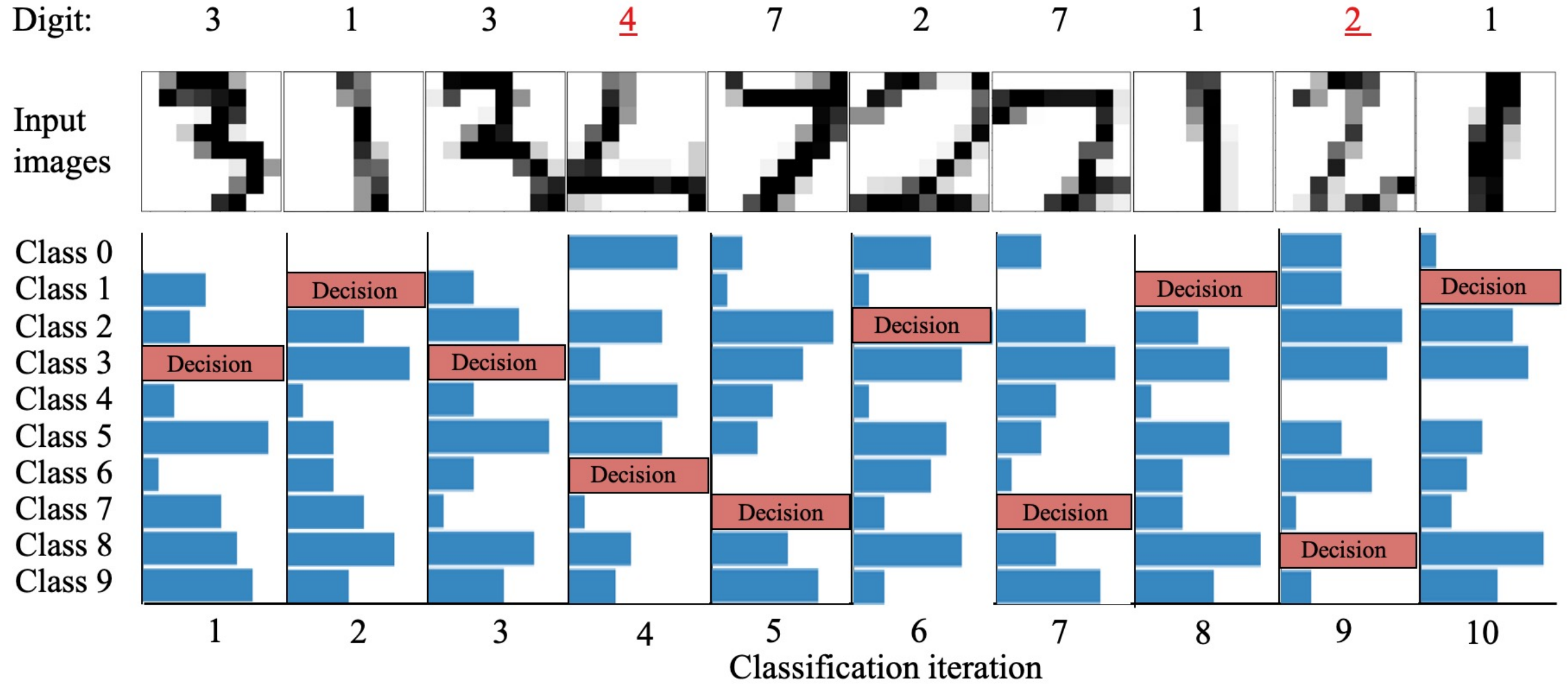} 
	\caption{Ten consecutive classifications of digits in MNIST dataset, as extracted based on SPICE simulations. For each classification, the height of the bars corresponds to the number of votes collected to each class. Note that total number of votes is 45 and equals to the total number of binary classifiers. Within a single classification period, the $i^\text{th}$ class, can get up to nine votes.}
	\label{fig:votes}
\end{figure*}
The subset of features is selected in a two-step approach: downsampling and feature selection. First, the features are uniformly downsampled from $28\times28$ pixels to $8\times 8$ pixels, as shown in Fig. \ref{fig:selection}. The downsampling of features significantly reduces the required hardware resources (\eg, 12 times less transistors is required) in exchange for 2.8\% accuracy degradation. A greedy feature selection algorithm, sequential backward selection (SBS) \cite{raschka2017python}, is used to select those most informative features (out of the remaining 64 features) for each binary classifier. As a result, the $8\times8$ features are reduced on average to 23 features per digit. Note that the number of selected features varies among the MNIST digits. For example, the digits 3 and 4 tend to look much more alike than the digits 0 and 1. Thus, significantly more features is selected for the 3-vs-4 binary classifier (44 features) than for the 0-vs-1 binary classifier (6 features). Comparing with other well-known feature selection algorithms (\eg, Fisher information \cite{guyon2003introduction}), SBS is determined to select the most informative features. Alternatively, SBS is an iterative greedy algorithm and is computationally expensive. For example, completing SBS on the original feature set of $28\times 28$ pixels requires 306,936 training iterations and 72 day (as extrapolated on shorter runs) on Intel Core i7-7700 CPU. Alternatively, with downsampled feature space only 2,016 training iterations which are completed within three hours on Intel Core i7-7700 CPU. The preferred set of features for each binary classifier is shown in Fig. \ref{fig:sbs}, exhibiting an average of 23 features per classifier.
\subsection{System Demonstration and Simulation Results} \label{sec:results}
The schematic representation of the overall system is shown in Fig. \ref{fig:overall_system}, comprising of AP-CNFET array to perform feature-weight products, voltage dividers to provide quantized features and feature weights, and buffers to extract the individual votes of each binary classifier.
The proposed system comprises 45 binary classifiers with a total of 1,021 AP-CNFETs utilized for the feature-weight products (shown by red and blue circles in Fig. \ref{fig:overall_system}). The product results are accumulated within the 45 sensing lines (one for each binary classifier). The system occupies \SI{3.8}{\square\micro\meter} as estimated based on transistor count and consumes \SI{295}{\pico\joule} energy per digit classification. Accuracy of 90\% is observed in SPICE on test set  of 10,000 unseen digits, as compared with the theoretical classification accuracy of 90\% obtained on the low resolution data set in Python. The confusion matrices obtained with Python and SPICE for the 10,000-digit test set are shown in Fig. \ref{fig:confusion_mat}. Note the similarities of the decisions extracted by SPICE and Python. The classifier is designed to operate at \SI{250}{\mega\hertz}, producing a single digit classification per cycle. The extracted votes and resultant decisions are shown in Fig. \ref{fig:votes} for ten consecutive classifications, as extracted from SPICE. Each binary classifier votes for a single class. Thus, the total number of votes equals to the number of binary classifiers. Final decision is made based on the class with highest number of votes within each classification period, as shown in the Fig. \ref{fig:votes}. Note the similarity of the incorrectly classified images (\textquoteleft4\textquoteright \space and \textquoteleft2\textquoteright) to the predicted labels (\textquoteleft6\textquoteright \space and \textquoteleft8\textquoteright, respectively).

Performance characteristics are listed in Table I for the proposed system along with the existing state-of-the-art conventional CMOS and emerging device memristor based classifiers \cite{wang2017low,krestinskaya2018learning}. For fair comparison, total current per decision (energy divided by supply voltage) and the system area normalized by squared form factor are also shown in Table I. The total current per decision with the proposed classifier is approximately 5.4 times lower as compared with state-of-the-art MOSFET approaches. Similarly, area savings range between $\times 15$ and $\times 859$ as compared to, respectively, MOSFET and memristor based classifiers, as shown in Table I. Alternatively, the proposed classifier exhibits only 2\% lower accuracy as compared with the artificial neural network (ANN) based classifiers in \cite{krestinskaya2018learning}.

\begin{figure}[t]
	\centering
	\includegraphics[width=1\columnwidth]{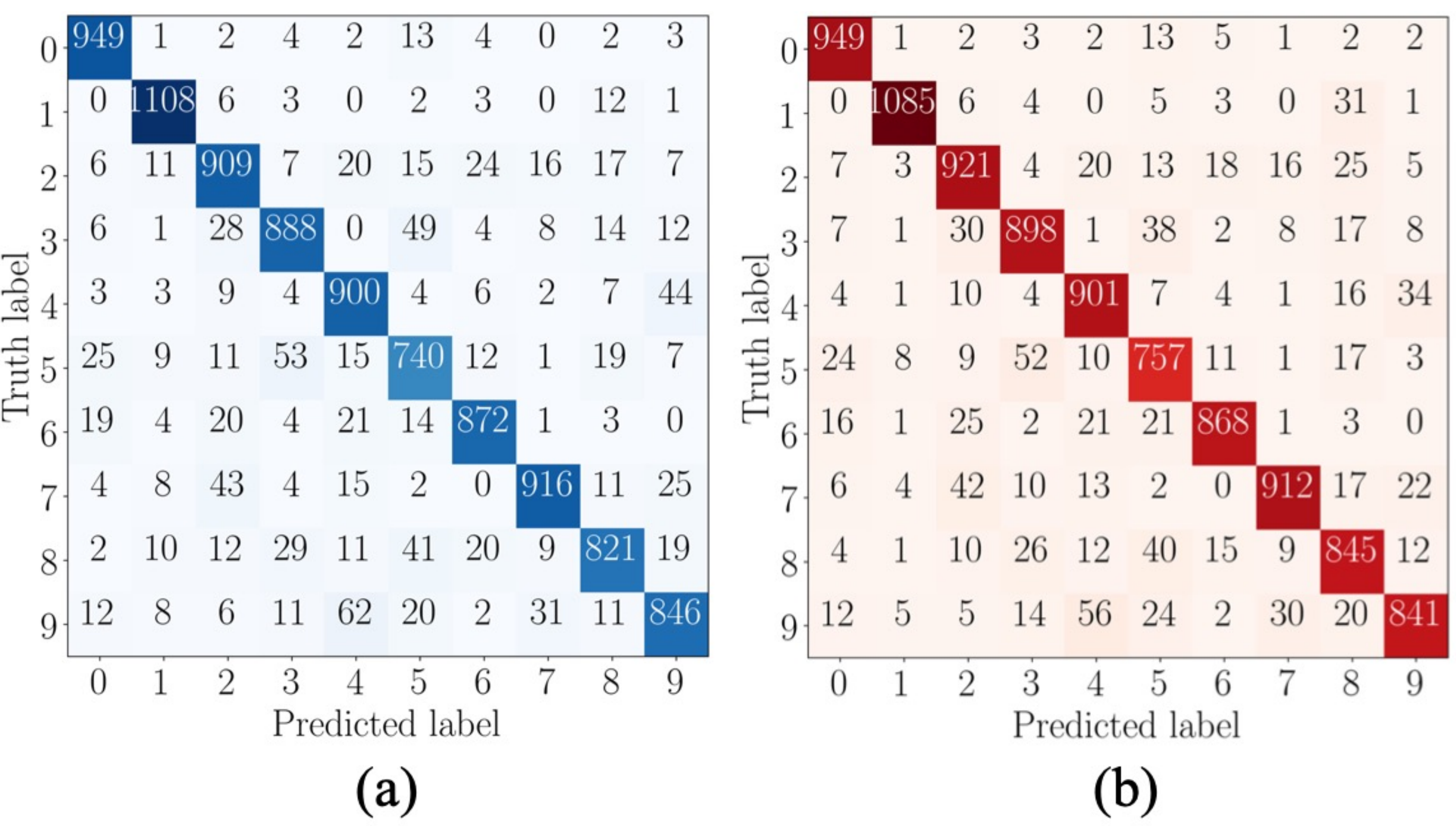} 
	\caption{Confusion matrices obtained by classifying MNIST in, (a) Python, and (b) SPICE.}
	\label{fig:confusion_mat}
\end{figure}
{\renewcommand{\arraystretch}{1.8}
	\begin{table}[t]
		\resizebox{\textwidth}{!}{%
			\renewcommand{\thetable}{\Roman{table}}
			
			\caption{\small{System characteristics of the proposed and a state-of-the-art CMOS and memristor based ML classifier.}}
			\vspace{-10pt}
			\label{table:evaluation}
			\centering
			\Huge	
			\begin{tabular}{|c|c|c|c|c|}				
				\hline	
				\multicolumn{2}{|c|}{} &\cite{krestinskaya2018learning}&\cite{wang2017low}&Current work\\ \hline
				\multicolumn{2}{|c|}{Device}& Memristor&MOSFET&AP-CNFET\\ \hline
				\multicolumn{2}{|c|}{Dataset}& MNIST&MNIST&MNIST\\ \hline
				\multicolumn{2}{|c|}{Technology}&\SI{180}{\nano\meter}&\SI{130}{\nano\meter}&\SI{15}{\nano\meter}\\ \hline
				\multicolumn{2}{|c|}{Algorithm} & ANN&Ada-boost&LR\\ \hline
				\multicolumn{2}{|c|}{Accuracy} & 92\%&90\%&90\%\\ \hline
				\multicolumn{2}{|c|}{\makecell{Offset from the ideal \\ accuracy}} & -1\%&0\%&0\%\\ \hline
				\multicolumn{2}{|c|}{Number of features} & 784&48&64\\ \hline		
				\multicolumn{2}{|c|}{Supply voltage} &\SI{1}{\volt}&\SI{1.2}{\volt}&\SI{3}{\volt} \\ \hline
				\multicolumn{2}{|c|}{Speed} &3.33 KHz&\SI{1.3}{\mega\hertz}&\SI{250}{\mega\hertz} \\ \hline
				\multirow{6}{*}{\begin{turn}{90}Costs\end{turn}}  
				&Energy&\SI{24}{\micro\joule}${}^{*}$& \SI{543}{\pico\joule}& \SI{295}{\pico\joule}${}^{**}$\\ 
				\cline{2-5}
				&Total current per decision& \SI{24}{\micro\ampere}$\cdot$sec& 453 pA$\cdot$sec& 98 pA$\cdot$sec\\ 
				\cline{2-5}				
				&Area&8,\SI{364}{\square\micro\meter}&246,\SI{792}{\square\micro\meter}&\SI{3.8}{\square\micro\meter}${}^{**}$\\ 
				\cline{2-5}	
				&Normalized area& \SI{0.26}{\micro\square\meter/\nano\square\meter}&\SI{14.6}{\micro\square\meter/\nano\square\meter}&\SI{0.017}{\micro\square\meter/\nano\square\meter}\\ 
				\cline{1-5}				
		
  \multicolumn{5}{l}{\Huge{$^{*}$ extrapolated based on the numbers reported in \cite{krestinskaya2018learning}  for a network with\vspace{-20pt}}}		\\    \multicolumn{5}{l}{\Huge{${}$ four input neurons and ten output neurons.\vspace{-20pt}}}	\\
    \multicolumn{5}{l}{\Huge{$^{**}$ overhead reported for MAC array. Note that the overheads of the system \vspace{-20pt}}}	\\   		
    \multicolumn{5}{l}{\Huge{${}$ is dominated by MAC array.\vspace{-20pt}}}	\\
			\end{tabular}
		}		
		\vspace{5pt}
\end{table}}
\section{CONCLUSIONS} \label{sec:Conc}
Increasing the device density and power efficiency has become challenging as the conventional CMOS scaling approaches its physical limits.
Alternatively, emerging devices, such as AP-CNFETs are inherently intelligent. The seamless mapping of the AI logic primitives onto AP-CNFET increases by orders of magnitude the embedded AI per transistor. 

To the best of the authors knowledge, the proposed system is the first to demonstrate ML classification with a single sensing line. To evaluate the system, a multi-class logistic classifier is designed in SPICE and demonstrated on MNIST dataset. The classifier uses 1,021 AP-CNFETs ($\times$17 reduction in the transistor count as compared with state-of-the-art CMOS based classifiers \cite{zhang2017memory,wang2017low}) and generates predictions at 250 MHz. The system exhibits \SI{295}{\pico\joule} energy consumption and occupies \SI{3.8}{\square\micro\meter} as estimated based on the transistor count in SPICE. With the proposed configuration, the reduced MNIST dataset is classified with no reduction in the overall prediction accuracy as compared with the theoretical Python results.

Theoretical bounds on classification accuracy are a strong function of the classification algorithm. In this paper, linear classification is demonstrated as a proof of concept of AP-CNFET  based AI. The theoretical accuracy with linear classifiers is however limited to 94\% using the default dataset of all the 784 MNIST features \cite{kaggle}. Higher accuracy ($>98\%$) can be achieved with more complex algorithms, such as non-linear support vector machines (SVM) and deep neural networks (DNNs). 
Similar to linear classifiers, the operation of these complex network is dominated by feature-weight multiplication and product accumulation. The proposed framework is expected to significantly increase the AI density, while reducing the power and area overheads in complex ML networks. 


\begin{thebibliography}{10}
	\providecommand{\url}[1]{#1}
	\csname url@samestyle\endcsname
	\providecommand{\newblock}{\relax}
	\providecommand{\bibinfo}[2]{#2}
	\providecommand{\BIBentrySTDinterwordspacing}{\spaceskip=0pt\relax}
	\providecommand{\BIBentryALTinterwordstretchfactor}{4}
	\providecommand{\BIBentryALTinterwordspacing}{\spaceskip=\fontdimen2\font plus
		\BIBentryALTinterwordstretchfactor\fontdimen3\font minus
		\fontdimen4\font\relax}
	\providecommand{\BIBforeignlanguage}[2]{{%
			\expandafter\ifx\csname l@#1\endcsname\relax
			\typeout{** WARNING: IEEEtran.bst: No hyphenation pattern has been}%
			\typeout{** loaded for the language `#1'. Using the pattern for}%
			\typeout{** the default language instead.}%
			\else
			\language=\csname l@#1\endcsname
			\fi
			#2}}
	\providecommand{\BIBdecl}{\relax}
	\BIBdecl
	
	\bibitem{lee2018unpu}
	J.~Lee~\text{et al}., ``{Unpu: An energy-efficient deep neural network
		accelerator with fully variable weight bit precision},'' \emph{IEEE J.
		Solid-State Circuits}, vol.~54, no.~1, pp. 173--185, Oct 2018.
	
	\bibitem{zhang2017memory}
	J.~Zhang, Z.~Wang, and N.~Verma, ``{In-memory computation of a machine-learning
		classifier in a standard 6T SRAM array},'' \emph{IEEE J. Solid-State
		Circuits}, vol.~52, no.~4, pp. 915--924, Apr 2017.
	
	\bibitem{wang2017low}
	Z.~Wang and N.~Verma, ``{A low-energy machine-learning classifier based on
		clocked comparators for direct inference on analog sensors},'' \emph{IEEE
		Trans. Circuits Syst. I, Reg. Papers}, vol.~64, no.~11, pp. 2954--2965, Jun
	2017.
	
	\bibitem{bankman2018always}
	D.~Bankman~\textit{et al}., ``{An always-on 3.8 $\mu$J/86\% CIFAR-10
		mixed-signal binary CNN processor with all memory on chip in 28nm CMOS},'' in
	\emph{IEEE Int. Solid-State Circuits Conf. (ISSCC) Dig. Tech. Papers}, Feb
	2018, pp. 222--224.
	
	\bibitem{kenarangi2019single}
	F.~Kenarangi and I.~Partin-Vaisband, ``{A single-MOSFET MAC for confidence and
		resolution (CORE) driven machine learning classification},'' \emph{arXiv
		preprint arXiv:1910.09597}, Oct 2019.
	
	\bibitem{kang2009chip}
	K.~Kang and T.~Shibata, ``{An on-chip-trainable gaussian-kernel analog support
		vector machine},'' \emph{IEEE Trans. Circuits Syst. I, Reg. Papers}, vol.~57,
	no.~7, pp. 1513--1524, Jul 2009.
	
	\bibitem{gonugondla2018variation}
	S.~K. Gonugondla, M.~Kang, and N.~R. Shanbhag, ``{A variation-tolerant
		in-memory machine learning classifier via on-chip training},'' \emph{IEEE J.
		Solid-State Circuits}, vol.~53, no.~11, pp. 3163--3173, Sep 2018.
	
	\bibitem{kang2018multi}
	M.~Kang~\text{et al}., ``{A multi-functional in-memory inference processor
		using a standard 6T SRAM array},'' \emph{IEEE J. Solid-State Circuits},
	vol.~53, no.~2, pp. 642--655, Jan 2018.
	
	\bibitem{amaravati201855}
	A.~Amaravati~\text{et al}., ``{A 55-nm, 1.0--0.4 v, 1.25-pj/mac time-domain
		mixed-signal neuromorphic accelerator with stochastic synapses for
		reinforcement learning in autonomous mobile robots},'' \emph{IEEE J.
		Solid-State Circuits}, vol.~54, no.~1, pp. 75--87, Dec 2018.
	
	\bibitem{hu2018memristor}
	M.~Hu~\text{et al}., ``{Memristor-based analog computation and neural network
		classification with a dot product engine},'' \emph{Advanced Materials},
	vol.~30, no.~9, p. 1705914, Jan 2018.
	
	\bibitem{yu2015scaling}
	S.~Yu~\text{et al}., ``{Scaling-up resistive synaptic arrays for neuro-inspired
		architecture: challenges and prospect},'' in \emph{Int. Electron Devices
		Meeting}, Dec 2015, pp. 17--3.
	
	\bibitem{agarwal2016resistive}
	S.~Agarwal~\text{et al}., ``{Resistive memory device requirements for a neural
		algorithm accelerator},'' in \emph{Int. Joint Conf. on Neural Networks}, Jul
	2016, pp. 929--938.
	
	\bibitem{krestinskaya2018learning}
	O.~Krestinskaya, K.~N. Salama, and A.~P. James, ``{Learning in memristive
		neural network architectures using analog backpropagation circuits},''
	\emph{IEEE Trans. Circuits Syst. I, Reg. Papers}, vol.~66, no.~2, pp.
	719--732, Sep 2018.
	
	\bibitem{xiang2015surface}
	D.~Xiang~\text{et al}., ``{Surface transfer doping induced effective modulation
		on ambipolar characteristics of few-layer black phosphorus},'' \emph{Nature
		communications}, vol.~6, p. 6485, Mar 2015.
	
	\bibitem{lin2005high}
	Y.-M. Lin~\text{et al}., ``{High-performance carbon nanotube field-effect
		transistor with tunable polarities},'' \emph{IEEE Trans. on Nanotech.},
	vol.~4, no.~5, pp. 481--489, Sep 2005.
	
	\bibitem{ben2011efficient}
	M.~H. Ben-Jamaa, K.~Mohanram, and G.~De~Micheli, ``{An efficient gate library
		for ambipolar CNTFET logic},'' \emph{IEEE Trans. Comput.-Aided Design Integr.
		Circuits Syst.}, vol.~30, no.~2, pp. 242--255, Feb 2011.
	
	\bibitem{das2013wse2}
	S.~Das and J.~Appenzeller, ``{WSe2 field effect transistors with enhanced
		ambipolar characteristics},'' \emph{Applied physics letters}, vol. 103,
	no.~10, p. 103501, Sep 2013.
	
	\bibitem{o2007cntfet}
	I.~O'Connor~\text{et al}., ``{CNTFET modeling and reconfigurable logic-circuit
		design},'' \emph{IEEE Trans. Circuits Syst. I, Reg. Papers}, vol.~54, no.~11,
	pp. 2365--2379, Nov 2007.
	
	\bibitem{kenarangi2019leveraging}
	F.~Kenarangi and I.~Partin-Vaisband, ``{Leveraging independent double-gate
		FinFET devices for machine learning classification},'' \emph{IEEE Trans.
		Circuits Syst. I, Reg. Papers}, vol.~PP, no.~99, pp. 1--12, Jul 2019.
	
	\bibitem{hu2017transient}
	X.~Hu and J.~S. Friedman, ``{Transient model with interchangeability for
		dual-gate ambipolar CNTFET logic design},'' in \emph{Int. Symp. on Nanoscale
		Arch. (NANOARCH)}, Oct 2017, pp. 61--66.
	
	\bibitem{hu2017closed}
	------, ``{Closed-form model for dual-gate ambipolar CNTFET circuit design},''
	in \emph{Int. Symp. on Circuits and Syst. (ISCAS)}, Sep 2017, pp. 1--4.
	
	\bibitem{nelder2004generalized}
	J.~A. Nelder and R.~J. Baker, ``Generalized linear models,'' \emph{Encyclopedia
		of statistical sciences}, vol.~4, Jul 2004.
	
	\bibitem{aly2005survey}
	M.~Aly, ``Survey on multiclass classification methods,'' \emph{Neural Netw},
	vol.~19, pp. 1--9, 2005.
	
	\bibitem{raschka2017python}
	S.~Raschka and V.~Mirjalili, \emph{{Python Machine Learning}}.\hskip 1em plus
	0.5em minus 0.4em\relax Birmingham, U.K.: Packt, 2017.
	
	\bibitem{guyon2003introduction}
	I.~Guyon and A.~Elisseeff, ``{An introduction to variable and feature
		selection},'' \emph{J. Mach. Learn. Res.}, vol.~3, pp. 1157--1182, Jan 2003.
	
	\bibitem{kaggle}
	\BIBentryALTinterwordspacing
	kaggle Inc. (2018) {Public Leaderboard: MNIST}. [Online]. Available:
	\url{https://www.kaggle.com/numbersareuseful/public-leaderboard-mnist}
	\BIBentrySTDinterwordspacing
	
\end{thebibliography}
\end{document}